\documentclass[aps,prl,twocolumn,showpacs]{revtex4}
\usepackage{graphicx}
\usepackage{bm}
\usepackage{amssymb}
\usepackage[dvips]{color}
\begin{document}
\title{Inertial particles driven by a telegraph noise}
\author{
G. Falkovich$^1$, S. Musacchio$^1$, L. Piterbarg$^2$, M.
Vucelja$^1$} \affiliation{$^1$Weizmann Institute of Science,
Rehovot 76100 Israel\\$^2$University of Southern California, Los
Angeles USA}

\begin{abstract}
We present a model for the Lagrangian dynamics of inertial
particles in a compressible flow, where fluid velocity gradients
are modelled by a telegraph noise. The model allows for an
analytic investigation of the role of time correlation of the flow
in the aggregation--disorder transition of inertial particle. The
dependence on Stokes and Kubo numbers of the Lyapunov exponent of
particle trajectories reveals the presence of a region in
parameter space $(St, Ku)$ where the leading Lyapunov exponent
changes sign, thus signaling the transition. The asymptotics of
short and long-correlated flows are discussed, as well as the
fluid-tracer limit.
\end{abstract}


\maketitle
\section{Introduction}
The spontaneous formation of clusters of particles suspended in
chaotic flows may originate from two different physical processes:
compressibility of the fluid flow and particle inertia. In the
first case particles are trapped in regions of ongoing
compression, while in the second case inertia causes their
ejection from vortical regions. The underlying link between these
two phenomena is manifested in the limit of weak inertia, in which
particle dynamics in incompressible flows can be approximated by
that of tracers in a weakly compressible velocity
field~\cite{M87,BFF01}.

Clustering processes have been extensively studied, both in the
case of compressible flows \cite{SO93,SE02,BGH04,BDES04} and
inertial particles \cite{BFF01,B03,FP04,DMOW05}. Fractal dimension
of the clusters, obtained from the ratios of Lyapunov exponents of
particle trajectories, has revealed a powerful tool to quantify
the intensity of the clustering \cite{SO93,B03}, and has motivated
further studies of Lyapunov exponents of inertial particles
\cite{DMOW05,BBBCMT06}.

Aggregation is an extreme form of clustering, which occurs when
the trajectories of different particles tend to point-like
clusters. In this situation the senior Lyapunov exponent of
particle trajectories become negative, signaling that nearby
particles do not separate exponentially as expected in a chaotic
flows, but their paths eventually coalesce. This phenomenon has
been described for fluid tracers in compressible flows
\cite{CKV98,FGV}, as well as for light particles in incompressible
flows \cite{B03}.

Surprisingly, it has been shown that in compressible flows, where
fluid trajectories coalesce, large enough inertia can induce a
transition from the strong clustering regime into a week
clustering one, where particle trajectories remains chaotic and
the senior Lyapunov exponent is positive
\cite{D85,WM03,MW04,MWDWL05}. Analytic results on this
aggregation--disorder transition have been obtained under the
assumption that the carrier flow is short-correlated in time.

It is therefore natural to ask how time correlations of real
turbulent flows can influence this phenomena. As pointed out by
recent numerical studies of fluid tracers in compressible flows
\cite{BDES04}, time correlations are responsible for an increase
in the level of compressibility required for the transition to the
strong-clustering regime. In the case of inertial particles in has
been found \cite{BBBCMT06} that time correlations cause increase
the (positive) Lyapunov exponent (i.e. make particle flow more
chaotic) already at weak inertia, which can not be predicted in
the framework of short correlated flows
\cite{DMOW05,BCH07,DFTT07,FH07}.

Here we discuss a simple time-correlated model flow,
recently introduced in~\cite{FM07},
which allows to get a deeper insight on the interplay between inertia
and compressibility, and to investigate analytically the dependence
on Stokes and Kubo numbers of Lyapunov exponent of particle trajectories.

\section{Model}

The dynamics of two small inertial particle, whose density is much
larger than the fluid density is dominated by the viscous drag.
Hence, the equation for their separation ${\bm R}(t)={\bm
X}_1(t)-{\bm X}_2(t)$ and relative velocity ${\bm V}$
reads~\cite{MR83}:
\begin{equation}
\dot{{\bm R}} = {\bm V} \;,\;\;
\dot{{\bm V}}= -\frac{1}{\tau}(\bm{V} - \Delta {\bm U}({\bm X_1},{\bm X_2},t))
\label{eq:1}
\end{equation}
where $\tau$ is the Stokes time of the particles,
and $\Delta {\bm U}$ is the difference of fluid
velocities at particle positions.

To investigate the behavior of Lyapunov exponent one has to
consider  separations  smaller than the viscous length-scale of
the flow. At these scales the fluid velocity difference can be
written in terms of the velocity gradients  $S_{ij} = \partial_j
U_i$ as $\Delta U_i = S_{ij}R_j$. The Lyapunov exponent $\lambda$
is determined by the contributions of the local gradients
$S_{ij}(t)$ experienced by the two nearby trajectories.
Note that  the statistics of velocity gradients in the reference
frame of an inertial particle differs from the Eulerian statistics
due to preferential concentration.

Let us now consider an idealized one-dimensional flow, in which
the fluid velocity gradient in the reference frame of a particle
is modelled by a {\it telegraph noise} $s(t)$, i.e. a noise that
switches randomly between two fixed values $\pm s$. Stokes number
can be defined in term of the fluid gradient intensity as $St = s
\tau$. Equations~(\ref{eq:1}) reduce to
\begin{equation}
\dot{R} =V \;,\;\;
\dot{V} = -\frac{V}{\tau}
+ \frac{s(t)}{\tau} R
\label{eq:system}
\end{equation}

When $s(t) =+s$ the system is expanding, and
evolves towards the asymptote $V= \alpha R$, where
\begin{equation}
\alpha = \frac{1}{2\tau}(-1+\sqrt{1+4 s \tau})\;.
\label{eq:alpha}
\end{equation}
The particles separate exponentially with smaller rate than
fluid trajectories $\alpha < s$.
When $s(t) =-s$ the system is contracting,
and two different scenarios are observed,
according to the relative intensity of fluid gradients
and inertial drag.
For small Stokes numbers $St <1/4$
the system evolves toward the asymptote $V= \beta R$, where
\begin{equation}
\beta = \frac{1}{2\tau}(-1+\sqrt{1-4 s \tau})\;\; for\;\; s\tau <1/4\;.
\label{eq:betaS}
\end{equation}
Particles slow down less efficiently than the fluid, and hence
their separation goes to zero with a faster exponential rate
$|\beta| > s$.
For large Stokes number  $St >1/4$
the system has two complex conjugate eigenvalues
\begin{equation}
\beta = \frac{1}{2\tau}(-1\pm i\sqrt{4 s \tau-1 })\;
for\;\; s\tau >1/4\;.
\label{eq:betaL}
\end{equation}
and the solution decays exponentially with a clockwise spiral
motion in phase space $(R,V)$. This means that particles can cross
the $R=0$ axis, i.e. collide, with non-zero relative velocity,
giving origin to a shock i.e. discontinuity in the velocity field
\cite{FFS02}. Notice that the solution is properly defined also
for $R<0$. The change of the sign of $R$ can be interpreted as the
fast particle overcoming the slow one.

Shocks start to appear for the critical value $St^* = 1/4$. The
presence of such critical Stokes number is characteristic of flows
where fluid velocity gradient are bounded, and its value is
determined by the intensity of the strongest negative gradient
(here $-s$). Conversely, if the statistics of fluid velocity
gradient is unbounded, shocks can appear for arbitrarily small
Stokes number, but they are exponentially suppressed in the limit
$St \to 0$ \cite{FFS02,WMB06,DFTT07}.

The transition rates $\nu_1$ from $-s$ to $s$ and $\nu_2$ from $s$
to $-s$ determine the fraction of time spent by the particle pair
in regions of ongoing compression and expansion (respectively,
$\nu_2/(\nu_1+\nu_2)$ and $\nu_1/(\nu_1+\nu_2)$). The mean value
of the noise $s(t)$ is
\begin{equation}
s_0 \equiv \langle s(t) \rangle = - s{\Delta \nu}/{\nu}\,,
\label{eq:s0}
\end{equation}
where $\nu = \nu_1+\nu_2$ and $\Delta \nu = \nu_2 -\nu_1$.
Noise fluctuations $\tilde{s}(t) \equiv s(t) - s_0$ are exponentially correlated:
\begin{equation}
\langle \tilde{s}(t) \tilde{s}(t') \rangle =
\frac{4s^2\nu_1\nu_2}{\nu^2}\exp(-\nu |t-t'|)
\label{eq:stilde}
\end{equation}
The ratio between the correlation time of the flow $1/\nu$ and the
Lagrangian separation time $1/s$ defines the Kubo number $Ku=
s\nu^{-1}$.

Notice that in potential flows both fluid tracers and inertial
particles spend more time in regions of a local compression than
in expanding regions, and hence $\Delta \nu >0$. The actual value
of $\Delta \nu$ can be determined by assuming statistical
homogeneity and isotropy of the flow. Thanks to these symmetries,
the mean position of a particle is solely determined by its
initial position $\langle X(t)\rangle = X_0$ and hence the
distance $R$ is statistically conserved.

The peculiar nature of the telegraph noise, namely the fact that
its square is deterministic, allows to obtain closed equations
for $\langle R(t)\rangle$. Averaging over different
realizations of the noise, and
using the derivation formula~\cite{SL78}
\begin{equation}
\frac{d}{dt} \langle \xi_t F_t(\xi) \rangle =
\langle \xi_t \frac{d}{dt} F_t(\xi) \rangle
-\nu \langle \xi_t F_t(\xi) \rangle
\label{eq:4}
\end{equation}
which hold for a generic noise with zero mean and correlation
$\langle \xi_t \xi_{t'} \rangle \sim \exp(-\nu |t-t'|)$,
one obtain the linear system
\begin{equation}
\frac{d}{dt} {\bm y} = \frac{1}{\tau} M {\bm y} \nonumber
\label{eq:systemMR}
\end{equation}
where
\begin{eqnarray}
{\bm y} &=&
\left[ \begin{array}{c}
\langle R \rangle \\
\tau \langle V \rangle \\
\tau \langle \tilde{s} R \rangle \\
\tau^2 \langle \tilde{s} V \rangle \\
\end{array} \right] \nonumber\\
M &=&
\left[ \begin{array}{cccc}
0 & 1 & 0 & 0 \\
s_0\tau & -1 & 1 & 0 \\
0 & 0 & -\nu \tau & 1 \\
(s^2-s_0^2)\tau^2 & 0 & -s_0\tau & -(1+\nu\tau) \\
\end{array} \right]
\label{eq:systemMR2}
\end{eqnarray}
This linear system allows for a constant solution
 $\langle R \rangle = R_0$ when
$s_0 \nu(1+\nu\tau) +s^2 =0$
which together with Eq.~(\ref{eq:s0}) gives
\begin{equation}
\Delta \nu = \frac{s}{1+\nu\tau}
\label{eq:deltanu}
\end{equation}

Notice that inertia reduces the trapping of particle in compressing regions.
This effect is encoded in the behavior of the parameter $\Delta \nu$,
which is a decreasing function of Stokes time $\tau$,
and vanishes in the limit $\tau \to \infty$.

\begin{figure} [t!]
\centerline{
\includegraphics[scale=0.75,draft=false]{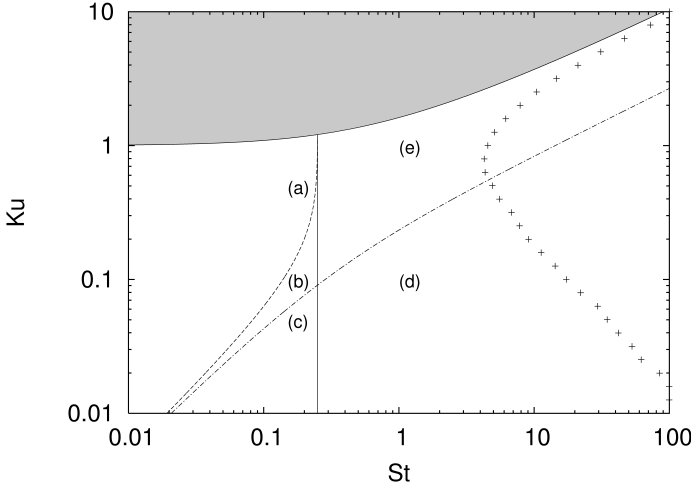}}
\caption{Parameter space $(St,Ku)$. Shape transitions in the
probability density function (pdf) of $x$ occur for $St=St^*=1/4$
(vertical solid line), $\tilde{m}=1$ (dashed line) and $m=1$
(dash-dotted line). Labels refers to the pdfs shown in
Fig.~\ref{pdfst0.125},~\ref{pdfst1}. Grey region is not accessible
by the model because $Ku$ number does not fulfill
condition~(\ref{eq:kumax}). Crosses represent the boundary of the
chaotic region.} \label{landscape}
\end{figure}

The positivity of $\nu_1$ requires $\Delta \nu <\nu$,
and gives an upper bound for the Kubo number achievable in the system:
\begin{equation}
Ku \leq Ku^* = \frac{2St}{\sqrt{1+4St}-1} \label{eq:kumax}
\end{equation}
For longer correlation time particle are irreversibly
captured by contracting regions.
In the limit of fluid tracers $\tau \to 0$ the constraint becomes $Ku<1$.
The accessible region in $(St,Ku)$ parameter space
is shown in Figure~\ref{landscape}.

\section{Statistics of particle-velocity gradients}
\label{sec:pdf}

From the system  (\ref{eq:system}) one can obtain a closed
equation for the particle-velocity gradient $\sigma = V/R$:
\begin{equation}
\dot{\sigma} = - \sigma^2 -{\tau}^{-1} \left[\sigma - s(t)
\right]\ . \label{eq:2}
\end{equation}
The Lyapunov exponent of particle trajectories can be obtained from
Eq.~(\ref{eq:2}) as an ensemble average $\langle \sigma \rangle$
over different realizations of the noise.
Equations for the stationary distribution of $x=\sigma + 1/(2\tau)$,
 are easily obtained
with the same procedure adopted
to derive Eq.~(\ref{eq:systemMR},\ref{eq:systemMR2}):
\begin{eqnarray}
\left[\left(\frac{1}{4\tau^2}-x^2+\frac{s_0}{\tau}\right)p\right]_x + q_x
&=& 0 \nonumber \\
\label{eq:5a}\\
\left[\left(\frac{1}{4\tau^2}-x^2-\frac{s_0}{\tau}\right)q\right]_x
+\left(\frac{s^2-s_0^2}{\tau^2}\right)p_x +\nu q&=& 0 \nonumber\\
\label{eq:5}
\end{eqnarray}
where 
$q(x)=\int \tilde{s} p(x,\tilde{s})d\tilde{s}/\tau$ and
$p(x,\tilde{s})$ is the joint stationary pdf of $x$ and
$\tilde{s}$. From the first equation one gets
$q=-(C+(1/4\tau^2-x^2+s_0/\tau)p)$ where $C$ is the mean flux of
$x$, and finally
\begin{eqnarray}
\left[\frac{s^2}{\tau^2}-\left(\frac{1}{4\tau^2}-x^2\right)^2\right]p_x
&+&\label{eq:6}\\
\left[(4x -\nu)\left(\frac{1}{4\tau^2}-x^2\right) -\nu
\frac{s_0}\tau \right] p &+& C(2x-\nu) = 0 \nonumber
\end{eqnarray}
\begin{figure} [t!]
\centerline{
\includegraphics[scale=0.75,draft=false]{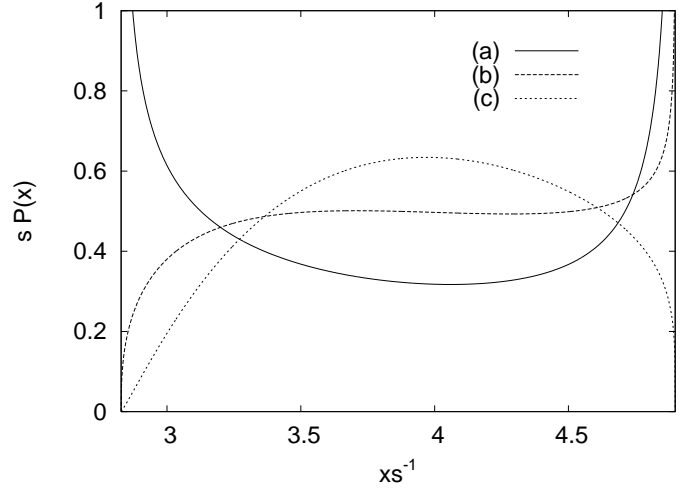}}
\caption{Pdf of $x$ in the small Stokes regime ($St=1/8$)
for different value of Kubo number:
$K=1/8 (a)$, $Ku=1/16 (b)$ and $Ku=1/24 (c)$.}
\label{pdfst0.125}
\end{figure}
The asymptotic behavior $p\sim C/x^2$ ($C>0$) for large $|x|$
gives the probability of strong particle velocity
gradients, and is therefore related to the probability of shocks.
Different solutions are found in the
small-Stokes-number and large-Stokes-number regime.

\subsection{Small Stokes number}
\label{sec:smallst}
When $St < St^*$ the unique positive integrable
solution of Eq.~(\ref{eq:6}) is obtained
under condition of zero flux ($C=0$):
\begin{equation}
p=C_1 \frac
{(w-x)^{m-1} (x-\tilde{w})^{\tilde{m}-1}}
{(w+x)^{m+1} (x+\tilde{w})^{\tilde{m}+1}}
\;,\;\;
x \in(\tilde{w},w)
\label{eq:7}
\end{equation}
and zero otherwise, where
\begin{equation}
\begin{array}{cc}
w=\sqrt{1+4s\tau}/2\tau & m=(\nu+\Delta\nu)/4w \\
\tilde{w}=\sqrt{1-4s\tau}/2\tau & \tilde{m}=(\nu-\Delta\nu)/4\tilde{w},
\label{eq:8}
\end{array}
\end{equation}

The solution is localized in the compact interval $(\tilde{w},w)$.
Its shape is determined by the values of $m$ and $\tilde{m}$. For
$\tilde{m}<1$ (low frequency) the pdf is peaked around the two
border values. For $m<1<\tilde{m}$ (intermediate frequency) it
vanishes at $\tilde{w}$, and finally when $m>1$ (high frequency)
it vanishes both at $\tilde{w}$ and $w$ (See
Fig.~\ref{pdfst0.125}). Indeed, when $St<St^*$, the solution of
the linear system for $(R,V)$ oscillates between two asymptotes
$V=(w-1/2\tau)R$ and $V=(\tilde{w}-1/2\tau)R$ according to the
sign of the noise. If the noise frequency is low the system has
enough time to get close to the two asymptotes and the pdf is
peaked around them. Conversely, when the sign of the noise
switches frequently, the system does not have enough time to reach
the asymptotes and oscillates rapidly around the mean value.

\subsection{Large Stokes number}
\label{sec:largest}
When $St >St^*$ the solution of Eq.~(\ref{eq:6}) consists of two
different parts:
\begin{equation}
p(x)=Cp_1(x)+C_2p_2(x)
\label{eq:11}
\end{equation}
The first one is the solution of Eq.~(\ref{eq:6}) with $C=1$,
\begin{eqnarray}
p_1 & = & \frac
{|w-x|^{m-1}}
{|w+x|^{m+1} (x^2+\tilde{w}^2)}
e^{n(\arctan(x/\tilde{w}))}
\label{eq:12}\\
& \times & \int_{-w}^x dy \frac {|w+y|^{m+1} (\nu-2y)}
{|y-w|^{m-1}(w^2-y^2)} e^{-n(\arctan(y/\tilde{w}))} \nonumber
\end{eqnarray}
while the second one is the right tail of the solution of the
homogeneous (fluxless) equation:
\begin{eqnarray}
p_2 & = & \frac
{|w-x|^{m-1}}
{|w+x|^{m+1} (x^2+\tilde{w}^2)} \nonumber \\
&\times & e^{n(\arctan(x/\tilde{w}))} I_{w,\infty}(x)\,.
\label{eq:13}
\end{eqnarray}

Here $I_{w,\infty}(x)$ is the characteristic function (indicator)
of the interval $(w,\infty)$, and
\begin{equation}
\begin{array}{cc}
w=\sqrt{4s\tau+1}/2\tau & m=(\nu+\Delta\nu)/4w \\
\tilde{w}=\sqrt{4s\tau-1}/2\tau & n=(\nu-\Delta\nu)/2\tilde{w}.
\label{eq:14}
\end{array}
\end{equation}
Conditions for determining $C$ and $C_2$ are
\begin{equation}
\int p(x)dx=1,\;\; \int q(x)dx=0\;.
\label{eq:15}
\end{equation}
These conditions guarantee the continuity of the pdf in the limit
$St\to St^*$ (see Appendix~\ref{appendixa})
\begin{figure}[t!]
\centerline{
\includegraphics[scale=0.75,draft=false]{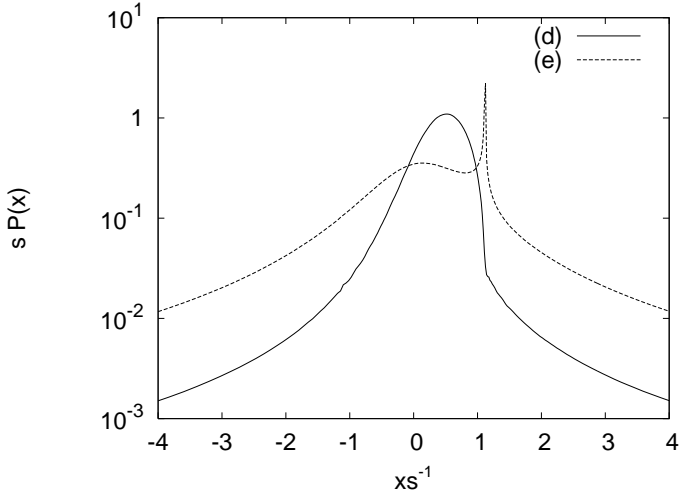}}
\caption{Pdf of $x$ in the large Stokes regime ($St=1$) for
for different values of Kubo number:
$Ku=0.1 (d)$, $Ku=1 (e)$}
\label{pdfst1}
\end{figure}
The pdf obtained in the regime $St>St^*$ is extended over all real
$x$, with power-law tails $p\sim C/x^2$ for large $|x|$ that are
due to shocks, which occur for large negative values of $s(t)$.
For short-correlated noise ($m>1$), the pdf is characterized by an
asymmetric core localized between $-w$ and $w$. When $m<1$ a
singular peak arises at $x=w$ (See Fig.~\ref{pdfst1}). This
behavior is easily understood in term of solution of the linear
system for $(R,V)$. When $s(t)=+s$ the solution converges toward
the asymptote $V=\alpha R =(w-1/2\tau)R$. This produces the peak
at $x=w$, provided that the correlation time of the noise is long
enough ($m<1$) to get close to the asymptote. The large-Stokes
number regime is hence characterized by an infinite series of
shocks alternated to 'quiet' phases in which the particle-velocity
gradient relaxes toward $\alpha$.

The schematic in Figure~\ref{landscape} summarizes the different regions
in the parameter space ($St$, $Ku$) where shape transitions occurs in the
pdf of $x$.

\section{Lyapunov exponents}
\label{sec:lyap}
\begin{figure} [t!]
\centerline{
\includegraphics[scale=0.75,draft=false]{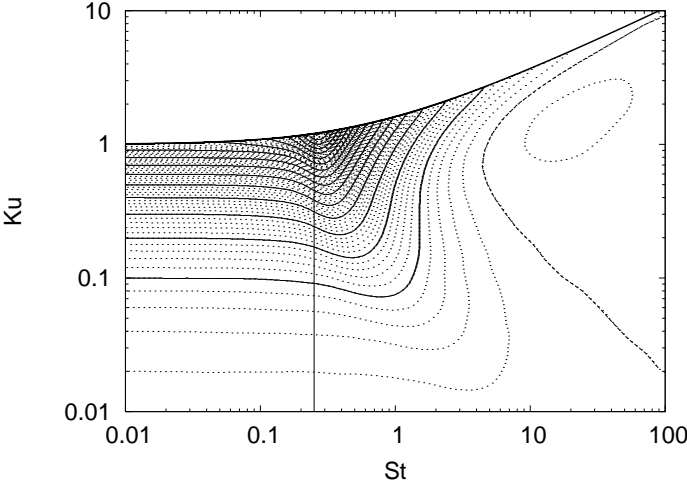}}
\caption{The isolines of the Lyapunov exponent $\lambda$ on the
plane of Stokes and Kubo numbers. The isolines are spaced every
$0.02s$ The boundary of the chaotic region ($\lambda >0$) is
represented by the dashed line.} \label{lyap}
\end{figure}

\begin{figure} [t!]
\centerline{
\includegraphics[scale=0.75,draft=false]{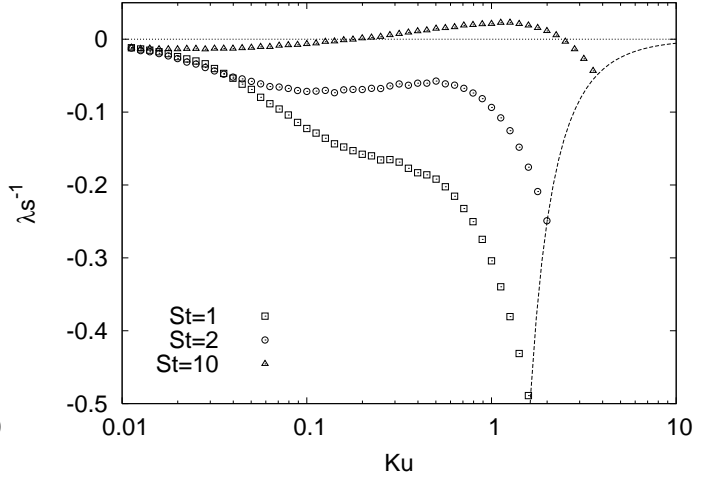}}
\caption{The Lyapunov exponent $\lambda$ as a function of the Kubo
number. An interval of positive Lyapunov exponents appears for $St
\gtrsim 4.3$. The dashed line represents the asymptotic behavior
for $Ku=Ku^*$.} \label{lyapku}
\end{figure}

The Lyapunov exponent of inertial-particle trajectories
can be written in terms of the mean value $\bar{x}$ as
$\lambda = \bar{x} -1/2\tau$.
For $St<St^*$ the mean value $\bar{x}$ can be written as
\begin{eqnarray}
\bar{x} & = & \tilde{w} \frac{(w-\tilde{w})\tilde{m}}{(m+\tilde{m})}
\nonumber \\
& \times & \frac{F_1(\tilde{m}+1,m+1,\tilde{m}+1,\tilde{m}+m+1,u,v)}
{F_1(\tilde{m}+,m+1,\tilde{m}+1,\tilde{m}+m,u,v)}
\label{eq:9}
\end{eqnarray}
where
\begin{equation}
u=\frac{\tilde{w}-w}{\tilde{w}+w},\;v=\frac{\tilde{w}-w}{2\tilde{w}}
\label{eq:10}
\end{equation}
and $F_1$ is the hypergeometric function of two variables\cite{gradry}.
The behavior of the Lyapunov exponent as a function of
Stokes and Kubo number is shown in Figure~\ref{lyap}.
Lyapunov exponent is negative for small Stokes, and
decreases approximatively as $\lambda s^{-1} \sim - Ku$
at increasing $Ku$ numbers.
In the limit $St \to 0$ it recovers the actual value for fluid tracers
\cite{FM07}:
\begin{equation}
\lambda_0 \equiv \lim_{\tau \to 0} \lambda = -s^2/\nu
\label{eq:l0}
\end{equation}
Notice that fluid tracers are always in the aggregation regime
within this model, as signaled by the negative value of
$\lambda_0$. A sharp negative minimum $\lambda=-2s$ is found for
$St=1/4$ at $Ku=(\sqrt{2}+1)/2$. It corresponds to the maximum
aggregation of particle. As the correlation time of the flow
decreases, the minimum becomes less pronounced, and it moves to
larger Stokes numbers.
A region of positive Lyapunov exponents, is present in the
parameter-space for $St \gtrsim 4.3$. The isoline of vanishing
Lyapunov exponent which border this region represent the
transition from the strong clustering regime ($\lambda <0$) to the
chaotic regime ($\lambda >0$). Indeed, Figure~\ref{lyapku} shows
that, as Stokes number increases, the interval of Kubo numbers
appears where the Lyapunov exponent grows, and eventually becomes
positive. This can be understood as follows: to achieve an
effective mixing the correlation time of the fluid gradients must
be long enough to provide substantial stretching of particle
trajectories, but not too long, to avoid particle segregation in
compressing regions. Therefore the chaotic region is confined in a
window of $Ku$ numbers between a lower and an upper bound
determined by stretching efficiency and particle trapping
respectively.

Let us now consider the behavior of the Lyapunov moments
$\gamma_n$, defined as $\langle R^n\rangle \sim \exp(\gamma_n t)$.
The evolution of $\langle R^n\rangle$, for $n$ positive integer,
is determined by a closed system of $2(n+1)$ equation:
\begin{eqnarray}
\frac{d}{dt}\langle R^{n-i}V^{i}\rangle &=&
(n-i)\langle R^{n-i-1}V^{i+1} \rangle \nonumber\\
&-& \frac{i}{\tau} \langle R^{n-i}V^{i} \rangle \nonumber\\
&+& \frac{i}{\tau}(\tilde{s}+s_0) \langle R^{n-i+1}V^{i-1} \rangle\\
\frac{d}{dt} \langle \tilde{s} R^{n-i}V^{i}\rangle &=&
(n-i)\langle \tilde{s} R^{n-i-1}V^{i+1} \rangle \nonumber\\
& -& \left(\frac{i}{\tau}+ \nu \right) \langle \tilde{s} R^{n-i}V^{i} \rangle
\nonumber\\
&-& \frac{i}{\tau}s_0 \langle \tilde{s} R^{n-i+1}V^{i-1} \rangle
\label{eq:moments1}
\end{eqnarray}
where $i = 0,1,\cdots,n$.
The $n-th$ Lyapunov moment $\gamma_n$ is
the largest solution of the $2(n+1)$-th equation:
\begin{equation}
\det
\left[ \begin{array}{cc}
{\bm A} & {\bm C}\\
{\bm D} & {\bm B}\\
\end{array} \right] =0\,,
\label{eq:moments2}
\end{equation}
${\bm A}$ and ${\bm B}$ are tri-diagonal matrices with the
elements
\begin{equation}
\begin{array}{ll}
A_{i,i} = \gamma_n +i/\tau  & i=1,n \\
A_{i,i-1} = -i s_0/\tau  & i=2,n \\
A_{i,i+1} =  i -n  & i=1,n-1 \\
B_{i,i} = \gamma_n +\nu +i/\tau  & i=1,n \\
B_{i,i-1} = i s_0/\tau  & i=2,n \\
B_{i,i+1} =  i -n  & i=1,n-1
\end{array}
\end{equation}
and ${\bm C}$ and ${\bm D}$ are sub-diagonal matrices with the
elements
$C_{i,i-1} = -i/\tau$ for $i=2,n $ and $D_{i,i-1} =
i(s_0^2-s^2)/\tau$ for $i=2,n$.
In the limit $St \to 0$ one recovers the Lyapunov moments of fluid
tracers \cite{FM07}:
\begin{equation}
\gamma_n = \sqrt{\left(\frac{\nu}{2}\right)^2+s^2(n^2-n)}-\frac{\nu}{2}
\label{eq:gamma_tracers}
\end{equation}
Notice that for fluid tracers the sign of the
separation $R$ is preserved, and hence the
exponents $\gamma_n$ coincide with
the exponents $\tilde{\gamma}_n$ defined by
$\langle |R|^n\rangle \sim \exp(\tilde{\gamma}_n t)$.
The asymptotic linear behavior of $\gamma_n$ for large n
is the hallmark of the presence of an upper bound
for velocity gradients.

\subsection{Short-correlated flows}
\label{sec:shortcorr}
\begin{figure} [t!]
\centerline{
\includegraphics[scale=0.75,draft=false]{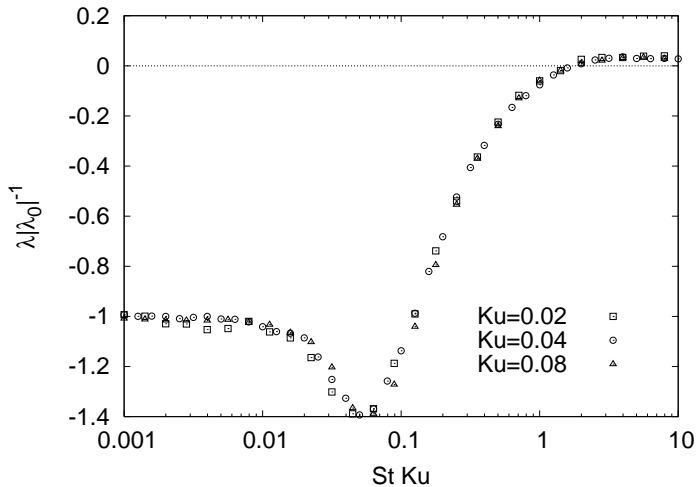}}
\caption{Lyapunov exponent $\lambda$ in the short-correlated limit.}
\label{lyapshort}
\end{figure}

Let us now examine the limit of short-correlated flow.
The limit $\nu \to \infty$ must be taken keeping
constant $s^2/\nu = D$, in order to recover
$\delta$-correlated noise fluctuations
$\langle s(t)s(t') \rangle = 2 D \delta(t-t')$
In other words, while Kubo number $Ku=s/\nu$ tends to zero,
Stokes number $St = s\tau$ must grow,
so that the product
\begin{equation}
Ku St = s^2/\nu \tau = D\tau
\end{equation}
remains constant. In this sense the short correlated limit for
inertial particles correspond always to the large-inertia case. In
the delta-correlated limit the relevant time-scale associated to
fluid velocity gradients is given by the Lyapunov exponent of
fluid tracers $\lambda_0 = - s^2/\nu =-D$. This is confirmed by
the collapse of particle Lyapunov exponents in the
short-correlated limit once their intensity and Stokes times are
re-scaled with $|\lambda_0|$ (see Fig.~\ref{lyapshort}).
A noticeable minimum is observed for
$|\lambda_0|\tau = Ku St \simeq 0.05$
and a transition to chaos, i.e.
from negative to positive $\lambda$, occurs for
$Ku St \gtrsim 1.6$. These features are in qualitative
agreement with previous analytic and theoretical results
obtained in the framework of $\delta$--correlated flows\cite{WM03,DFTT07}.
Notice that in those studies Gaussian statistics is assumed for velocity gradients,
at variance with our model in which only the two values $\pm s$
are allowed. Therefore quantitative details such as the exact
position of the minimum can be different.

We remark that in the short--correlated asymptotics, the
fluid--tracers limit becomes singular. Indeed one has
\begin{equation}
\lim_{\nu \to \infty} s_0 = \lim_{\nu \to \infty}
-\frac{s^2}{\nu(1+\nu\tau)}=
\left\{ \begin{array}{cc}
-D & \tau = 0 \\
0 & \tau >0 \\
\end{array} \right.
\end{equation}
which signals that fluid tracers are still preferentially attracted
by regions of ongoing compression, while inertial particles
with arbitrary finite $\tau$ are not.
Notice that in the limit $\nu \to \infty$
fluid velocity gradients becomes unbound because $s \to \infty$,
and for fluid tracers we recover the quadratic behavior
of Lyapunov moments $\gamma_n = D (n^2-n)$.

\subsection{Long-correlated flows}
\label{sec:longcorr}
\begin{figure} [t!]
\centerline{
\includegraphics[scale=0.75,draft=false]{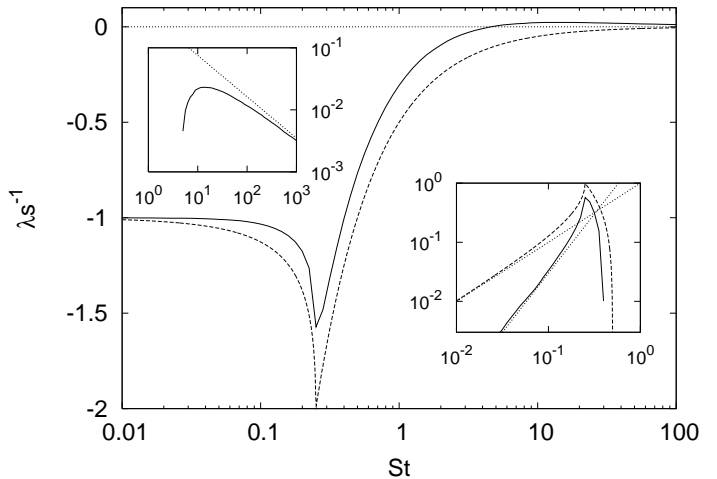}}
\caption{Lyapunov exponent $\lambda$ for $Ku=1$ (solid line)
and on the line $Ku=Ku^*$ (dashed line), corresponding to the
longest correlation of velocity gradients
achievable within the model.
Upper inset: asymptotic behavior for large Stokes number
$\lambda s^{-1} \sim St^{-2./3.}$, ($Ku=1$).
Lower inset: asymptotic behavior for small Stokes number
$|\lambda -\lambda_0|s^{-1} \sim St^{\zeta}$.}
\label{lyap_max}
\end{figure}

Time correlation of fluid gradient is bounded by the
condition~(\ref{eq:kumax}).
When $Ku=Ku^*$ the transition rate $\nu_1$ from $-s$ to $s$ vanishes,
while the transition rate $\nu_2$ from $s$ to $-s$ reaches its maximum
$\nu^*=(\sqrt{1+4 s \tau }-1)/(2 \tau) $.
Particles initially seeded in expanding regions
are gradually captured by contracting ones, where
they remain trapped forever. Therefore
the population of expanding regions decreases exponentially as
\begin{equation}
P(t) \sim \exp(-\nu^*t).
\label{eq:prob}
\end{equation}
Moments of particle separations will hence evolve
asymptotically according to
\begin{equation}
\langle |R|^n \rangle  \sim e^{n\alpha t}P(t) +
(1-P(t))e^{n Re(\beta) t}f(Im(\beta)t)
\label{eq:erren}
\end{equation}
where  $\beta$ is given by (\ref{eq:betaL}) and $f(t)$ is $2\pi$-
periodic function. Notice that $\nu^*= \alpha$ and therefore for
$n=1$ the decreasing fraction of particle in expanding regions is
exactly balanced by the exponential growth of their separation.
From (\ref{eq:prob},\ref{eq:erren}) one gets the Lyapunov moments
\begin{equation}
\begin{array}{cc}
\tilde{\gamma}_n = \alpha(n-1) & for\; n \ge (\alpha-Re(\beta))^{-1} \\
\tilde{\gamma}_n = Re(\beta) n &  for\; n \le (\alpha-Re(\beta))^{-1} \\
\end{array}
\label{eq:gamman}
\end{equation}

In the limit of fluid tracers one gets
\begin{equation}
\begin{array}{cc}
\tilde{\gamma}_n = s(n-1) & for\; n \ge 1/2 \\
\tilde{\gamma}_n = -s n &  for\; n \le 1/2 \\
\end{array}
\label{eq:gamma_tracers_kumax}
\end{equation}
in agreement with Eq.~(\ref{eq:gamma_tracers}).

Finally, the Lyapunov exponent along the critical line $Ku=Ku^*$ is
$\lambda = (\partial \tilde{\gamma}_n /\partial )\arrowvert_{n=0} = Re(\beta)$
In Figure~\ref{lyap_max} we compare its behavior with that
of the Lyapunov exponent along the line $Ku=1$.
For small Stokes number both of them recover the fluid-tracers limit
$\lambda \to \lambda_0 = -s$ but with different power law behavior.
On the line $Ku=Ku^*$ we have $|\lambda -\lambda_0|s^{-1} \sim St$,
while for $Ku=1$ we have $|\lambda -\lambda_0|s^{-1} \sim St^2$
(see lower inset of Fig.~\ref{lyap_max}).
For large Stokes number
Lyapunov vanishes as $\lambda s^{-1} \sim - St^{-1}$
on the line $Ku=Ku^*$ and as $\lambda s^{-1} \sim St^{-2/3}$ for $Ku=1$
(see upper inset of Fig.~\ref{lyap_max}).
In between these two asymptotics a sharp minimum appears for $St=St^*$.

Notice that the asymptotic decay $St^{-2/3}$ here shown for
long--correlated flows, have been already predicted and observed also
for $\delta$--correlated flows \cite{BCH07,FH07}.
The agreement between these results confirms that in the
large--Stokes number asymptotics the role of time correlation
becomes negligible and particles behave as if suspended in
$\delta$--correlated--in--time flows.

\section{Conclusions}
\label{sec:conclusion}

We discussed the Lagrangian dynamics of inertial particles in a
simple time-correlated compressible flow, in which fluid velocity
gradients in the reference frame of the particle are modelled by a
one-dimensional telegraph noise. In spite of its simplicity, the
model allows to take into account consistently the different
Lagrangian weights of region of ongoing compression and expansion,
and it reproduces the phenomenon of trapping of particles in
compressing regions for long-correlated flows.

The peculiar nature of the telegraph noise allows to
investigate analytically the effects of time-correlation
of velocity gradients on the
chaoticity of particle trajectories, and to study the dependence
on Stokes and Kubo numbers of Lyapunov exponents.
We discussed both the asymptotics of long and short-correlated flows,
as well as the fluid-tracers limit.

For large Stokes number, a regime characterized by the formation
of shocks, we found a chaotic region in parameter space
($St$,$Ku$), where the leading Lyapunov exponent becomes positive.
Inertia is therefore responsible for a transition from a strong
clustering regime, originated by the compressible nature of the
flow, to a chaotic regime. The latter is observed in a range of
$Ku$ numbers such that the time correlation of fluid gradients is
long enough to provide substantial stretching, but not too long to
get particles trapped in compressing regions.


\section{Acknowledgments}
This work has been supported by the ONR Grant N00014-99-0042 and
by the grant of the Israeli Science Foundation.
\section{Appendix A}
\label{appendixa}
To prove the continuity of the pdf for $St=St^*$
we first notice that the limit $St \to St^*$ is equivalent to
$\tilde{w} \to 0$.

The limit $\tilde{w} \to 0$ of the solution~(\ref{eq:7})
in the small-Stokes number regime is
$p(x) = C_1 f(x) I_{0,w}(x)$, where
\begin{equation}
f(x) =\frac
{(w-x)^{m-1}}
{(w+x)^{m+1} x^2}e^{-(\nu-\Delta\nu)/2x}
\label{eq:16}
\end{equation}
and $I_{0,w}(x)$ is the characteristic
function of the interval $(0,w)$.

To study the limit of the solution~(\ref{eq:11})
let us consider the limits of $p_1(x)$ and $p_2(x)$ when
$\tilde{w} \to 0$.
The latter is easily obtained as:
\begin{equation}
p_2(x) \sim e^{(\nu-\Delta\nu)\pi/4\tilde{w}}f(x)
\label{eq:17}
\end{equation}
Now let us rewrite $p_1(x)$ as
\begin{eqnarray}
p_1 & = & \frac
{|w-x|^{m-1}}
{|w+x|^{m+1} (x^2+\tilde{w}^2)}
e^{-n(\arctan(\tilde{w}/x))}
\nonumber \\
& \times &
\int_{-w}^x \frac
{|w+y|^{m+1} (\nu-2y)dy}
{|y-w|^{m-1}(w^2-y^2)}
e^{n(\arctan(\tilde{w}/y))}
\label{eq:18}
\end{eqnarray}
As $\tilde{w} \to 0$ then $p_1(x) = O(1)$ if $x<0$ and
\begin{equation}
p_1(x) \sim 2e^{(\nu-\Delta\nu) \pi/4\tilde{w}} \tilde{w}^2 f(x)
\label{eq:19}
\end{equation}
for $x>0$. To get the last asymptotic one takes into account
that the main contribution in the integral in~(\ref{eq:18})
is brought by the right small vicinity of $y=0$, say $(0,s)$,
and apply
\begin{equation}
\int_0^s \exp(-\arctan(y/\delta)/\delta)dy \sim
\int_0^s \exp(-y/\delta^2)dy \sim \delta^2.
\label{eq:20}
\end{equation}
for all $s>0$ and small $\delta$.

The continuity of the solution of Eq.~(\ref{eq:6}) in the
limit $\tilde{w} \to 0$ is therefore guaranteed by the
conditions
\begin{equation}
C=\frac{C_1}{2\tilde{w}^2e^{(\nu-\Delta\nu)\pi/4\tilde{w}}}, \;\;
C_2 = 2C\tilde{w}^2
\label{eq:22}
\end{equation}
This conditions are indeed equivalent to the normalizations
conditions
\begin{equation}
\int p(x)dx=1,\;\; \int q(x)dx=0
\label{eq:23}
\end{equation}
where $q=-(C+(1/4\tau^2+s_0/\tau-x^2)p)$.

To show this we notice that Eq.~(\ref{eq:6}) can be
written as
\begin{equation}
\left(\left(\frac{s^2}{\tau^2}-\left(\frac{1}{4\tau^2}-x^2 \right)^2 \right)p
+ Cx^2\right)_x +\nu q =0
\label{eq:24a}
\end{equation}
It follows that
\begin{equation}
\int_{-x}^x q(y)dy =-\frac {1}{\nu}
\left( \frac{s^2}{\tau^2}-\left(\frac{1}{4\tau^2}-x^2\right)^2\right)
(p(x)-p(-x))
\label{eq:24}
\end{equation}
Thus, the second condition in~(\ref{eq:23}) is equivalent to
\begin{equation}
\lim_{x\to \infty}x^4(p(x)-p(-x))=0
\label{eq:25}
\end{equation}
According to Eq.~(\ref{eq:11}) $p(x)$
is written as the sum of $p_1$ and $p_2$, whose asymptotic behavior is
\begin{equation}
p_1(x)\sim \frac{1}{x^2}+\frac{P_1^+}{x^4}, \ x\to \infty, \quad p_1(x)\sim \frac{1}{x^2}+\frac{P_1^-}{x^4}, \ x\to -\infty
\label{eq:26}
\end{equation}
and
\begin{equation}
p_2(x)\sim \frac{P_2^+}{x^4}, \ x\to \infty, \quad p_2(x)\sim \frac{P_2^-}{x^4}, \ x\to -\infty
\label{eq:27}
\end{equation}
Notice that $P_2^-=0, \ P_2^+=e^{(\nu-\Delta\nu)\pi/4\tilde w}$.
Thus~(\ref{eq:25}) becomes
\begin{equation}
C_2=\frac{C(P_1^+-P_1^-)}{P_2^+}
\label{eq:28}
\end{equation}
In the limit $\tilde{w} \to 0$
\begin{equation}
P_1^-=o(P_1^+), \ P_1^+ \sim 2\tilde w^2P_2^+
\label{eq:29}
\end{equation}
and we get
\begin{equation}
C_2\sim 2\tilde w^2C
\label{eq:30}
\end{equation}
which implies the continuity of the pdf.
Normalization of $p(x)$ in the limit  $\tilde{w} \to 0$
is equivalent to the first condition in~(\ref{eq:22}).



\end{document}